\documentclass[num-refs]{wiley-article}
\usepackage[numbers]{natbib}

\graphicspath{ {./figures/} }
\usepackage[hidelinks]{hyperref}
\usepackage{cleveref}
\usepackage{float}
\usepackage{verbatim} 
\usepackage{xcolor}
\usepackage{multirow}
\usepackage{booktabs}
\restylefloat{figure}
\restylefloat{table}
\usepackage{natbib}

\usepackage[autostyle, english = american]{csquotes}
\MakeOuterQuote{"}

\usepackage{siunitx}

\papertype{Original Article}
\paperfield{Journal Section}

\title{A Hybrid AI Methodology for Generating Ontologies of Research Topics from Scientific Paper Corpora}

\author[1]{Alessia Pisu}
\author[1]{Livio Pompianu}
\author[2,3]{Francesco~Osborne}
\author[1]{Diego Reforgiato Recupero}
\author[1]{Daniele~Riboni}
\author[2]{Angelo Salatino}

\affil[1]{Department of Mathematics and Computer Science, University of Cagliari, Cagliari, Italy}
\affil[2]{Knowledge Media Institute, The Open University, Milton Keynes, United Kingdom}
\affil[3]{Department of Business and Law, University of Milano Bicocca, Milano, Italy}

\corraddress{Diego Reforgiato Recupero}
\corremail{diego.reforgiato@unica.it}

\fundinginfo{This work was supported by multiple funding sources: 1) National Recovery and Resilience Plan (NRRP) under Mission 4, Component 2, Investment 1.5 (Call No. 3277, December 30, 2021), funded by the European Union – NextGenerationEU, for the project eINS – Ecosystem of Innovation for Next Generation Sardinia (Project Code: ECS0000038, CUP: F53C22000430001, Grant Decree No. 1056, June 23, 2022). 2) EU-FSE support under PON Research and Innovation 2014–2020 (D.M. 1061/2021) for the FAIR – Future Artificial Intelligent Research project (CUP: J53C22003010006) and the ANSWER cascade project (CUP: J23C24000080007). 3) FIATLUCS project, funded by PNRR RAISE Liguria, Spoke 01 (CUP: F23C24000240006).}

\runningauthor{Pisu et al.}

\begin{document}

\begin{frontmatter}
\maketitle

\begin{abstract}
Modern AI systems offer tools for literature search, automated reviews, hypothesis generation, and more. However, managing the increasing volume of academic publications produced each year continues to pose a significant challenge. Large Language Models (LLMs) 
have revolutionised text processing, but struggle with synthesising and navigating the structure of entire research fields.
To address this challenge, researchers have proposed developing structured, interlinked, and formal representations of the content of research publications, enabling more effective ingestion by AI systems. Among the various dimensions for describing the content of a research paper, the most fundamental is the research topic. Therefore, taxonomies and ontologies of research topics (e.g., MeSH, UMLS, CSO, NLM) play a central role in providing the primary framework through which intelligent systems can explore and interpret the literature.
However, these resources have traditionally been manually curated, a process that is time-consuming, prone to obsolescence, and limited in granularity. 
This paper presents Sci-OG, a semi-auto\-mated methodology for generating research topic ontologies, employing a multi-step approach: 1) Topic Discovery, extracting potential topics from research papers; 2) Relationship Classification, determining semantic relationships between topic pairs; and 3) Ontology Construction, refining and organizing topics into a structured ontology. 
The relationship classification component, which constitutes the core of the system, integrates an encoder-based language model with features describing topic occurrence in the scientific literature. We evaluate this approach against a range of alternative solutions using a dataset of 21,649 manually annotated semantic triples. Our method achieves the highest F1 score (0.951), surpassing various competing approaches, including a fine-tuned SciBERT model and several LLM baselines, such as the fine-tuned GPT4-mini. 
Our work is corroborated by a use case which illustrates the practical application of our system to extend the CSO ontology in the area of cybersecurity.
The presented solution is designed to improve the accessibility, organization, and analysis of scientific knowledge, thereby supporting advancements in AI-enabled literature management and research exploration.




\end{abstract}
\end{frontmatter}

\section{Introduction}
\label{sec:introduction}

We are at a pivotal moment in the history of science, characterized by the emergence of sophisticated artificial intelligence models designed to support scientific research and increased access to scientific publications through open access. These advancements have the potential to revolutionize scientific research and the manner in which scientists engage with and contribute to the collective scientific knowledge. The AI community is actively developing advanced techniques for effective literature search, navigation, and comprehension, as well as pioneering methods for the automated generation of literature reviews, improvement of academic writing and citation practices, automatic hypothesis generation, and more.

However, managing the vast and continuously growing body of scientific literature remains a significant challenge~\citep{bornmann2015,maslej2025artificialintelligenceindexreport}. While the introduction of Large Language Models (LLMs) has revolutionized the field of natural language processing (NLP)~\citep{kung2023performance,openai2023gpt4}, these models have limitations when it comes to processing long texts. Efficiently searching and summarizing this huge volume of documents remains difficult even with modern solutions such as Retrieval-Augmented Generation (RAG)~\citep{lewis2020retrieval}. Consequently, although these advanced models can address questions about specific papers, they struggle to comprehend the overall structure of a research field represented by millions of papers.

To tackle this issue, it was proposed to develop structured, interlinked, and formal representations of the content of research publications, which could be more easily ingested by AI systems~\citep{auer2018towards,kuhn2017genuine}.   
The most fundamental dimension for describing the concepts within a research paper, and thereby enabling a more comprehensive analysis of the literature, is the research topic~\citep{mazzocchi2018}. A research topic is typically defined as the subject of study or the issue of interest to the academic community, and it is explicitly addressed by research papers~\citep{salatino2019early}. Research topics, in various forms, are commonly represented in many repositories of article metadata. Indeed, a recent survey~\citep{salatino2025survey} identified 45 classification schemes for describing research topics. While a few of these adopt a shallow approach and represent topics as a flat list of keywords, most schemes provide greater value when the topics are organized into taxonomies or, more recently, as formal ontologies.
These taxonomies of research topics (e.g., MeSH, UMLS, CSO, NLM) play a crucial role in categorizing, managing, and querying scholarly information and serve as the primary framework for intelligent systems to explore and understand the literature. 
This includes search engine~\citep{gusenbauer2020academic}, conversational agents~\citep{meloni2023integrating}, analytics dashboards~\citep{angioni2022aida}, academic recommender systems~\citep{beel2016paper}, and many other tools in this space. A solid representation of research topics is also the foundation for many AI-driven literature analyses~\citep{goodell2021artificial,salatino2023diversity}.


Traditionally, taxonomies of research topics have been created manually. This process typically involves multiple experts agreeing on a general structure and filling in various branches.  However, this is an expensive and time-consuming process. Hence, by the time these taxonomies are released, they tend to be already obsolete, especially in fields such as Computer Science, where the most interesting topics are the newly emerging ones~\citep{osborne2012mining}. Moreover, these taxonomies are typically coarse-grained and usually represent wide categories of approaches rather than the fine-grained topics addressed by researchers. 

An alternative approach involves developing methods for the automatic or semi-automatic creation of research topic ontologies~\citep{osborne2015klink,osborne2018pragmatic}. 
A notable example of this approach is Klink-2~\citep{osborne2015klink}, which was developed by some of the authors of this paper with the aim of producing the Computer Science Ontology (CSO)~\citep{10.1162/dint_a_00055}. CSO has become one of the largest resources in the field, comprising about 14K topics and 159K semantic relationships, and it has been adopted by several organizations worldwide, including Springer Nature~\citep{osborne2016automatic}. 
Over the past two years, LLMs have significantly transformed the field of NLP and have been increasingly applied to ontology generation and matching. However, generating an ontology of research topics remains a challenging task even for these systems, as it requires information that can only be derived from large corpora of research articles. In particular, it involves identifying which new topics are emerging and determining their relationships with other topics, as perceived by domain experts authoring novel research.

This paper presents Sci-OG (Scientific Ontology Generation), a novel AI methodology for the semi-automatic generation of taxonomies of research topics that follows a multi-step process: 1) \textit{Topic Discovery}, which extracts potential topics from research papers; 2) \textit{Relationship Classification}, which determines the semantic relationships between pairs of topics; and 3) \textit{Ontology Construction}, which refines and organizes the topics into a structured ontology.

The relationship classification component, which forms the core of the system, integrates an encoder-based language model fine-tuned on high-quality data with features describing topic occurrence in the scientific literature. These features are extracted from a large-scale corpus of papers, allowing the component to exploit the implicit knowledge embedded in the corpus.
We validated the approach on \textit{CSO-21K}, a novel dataset covering 21,649 relationships, achieving excellent performance.  
The hybrid method proposed in this paper achieved the highest F1 score (0.951), outperforming both the fine-tuned SciBERT model and several LLMs, including the fine-tuned GPT4-mini.  

The novel approach was developed as the successor to Klink-2, with the goal of further expanding and refining the CSO. For this reason, this paper also presents a case study that applies this technology to enhance the CSO by developing a new branch focused on cybersecurity. 

In summary, the contributions of this paper are as follows:
\begin{itemize}
\item We propose a new AI methodology for the semi-automatic generation of research topic taxonomies.
\item We introduce \textit{CSO-21K}, a novel gold standard for research topic relation prediction, designed to support the evaluation of models in this domain.
\item We conduct an evaluation comparing our methodology with several alternatives, including pipelines based on LLMs.
\item We release the \textit{CSO-21K} benchmark, the evaluation data, and the models in an open repository\footnote{Repository - \url{https://github.com/aleessiap/scientific_ontology.git}}.
\end{itemize}

The remainder of this paper is structured as follows. Section~\ref{sec:relatedwork} reviews the relevant literature. Section~\ref{sec:background} describes the background data used in the experiments and details the gold standard.
Section~\ref{sec:method} presents the proposed methodology.
Section~\ref{sec:evaluation} reports the evaluation results. 
Section~\ref{sec:casestudy} discusses a case study in which we applied the proposed method to generate a novel cybersecurity ontology, which was then integrated into CSO. 
Finally, Section~\ref{sec:conclusions} concludes the paper and outlines directions for future work.

\section{Related Work}
\label{sec:relatedwork}


In this section, we first outline the main research taxonomies used in the computer science domain and then discuss the AI methodologies for the  generation of research taxonomies.



\subsection{Taxonomies in Computer Science}\label{sec:ontologies}

The field of Computer Science is covered by several research taxonomies.

The ACM Computing Classification System\footnote{The ACM Computing Classification System – \url{http://www.acm.org/publications/class-2012}} serves as a well-known taxonomy for categorizing research topics. This system, created and maintained by the Association for Computing Machinery (ACM), the largest computing society dedicated to education and science worldwide, includes about 2,000 research topics. 
However, the ACM Computing Classification System relies on manual curation, which makes updating it both time-consuming and expensive. Consequently, updates are rare, with the latest revision made in 2012, resulting in the taxonomy becoming outdated relatively quickly.


The CSO is one of the most comprehensive topic classifications, covering 14,000 research areas~\citep{salatino2018computer}. It was automatically generated using the Klink-2 algorithm~\citep{osborne2015klink} on a dataset of 16 million scientific articles. The CSO stands out due to two main benefits: i) it provides an exceptionally detailed and nuanced representation of the field, and ii) it can be easily updated by applying Klink-2 to new publication datasets. CSO is integral to several tools used by Springer Nature's editorial team, aiding in research publication classification, identification of research communities, and prediction of research trends~\citep{10.1162/dint_a_00055}.




The IEEE Taxonomy primarily covers the field of Engineering but also includes various concepts relevant to Computer Science. Developed and maintained by the Institute of Electrical and Electronics Engineers\footnote{IEEE Taxonomy - \url{https://www.ieee.org/content/dam/ieee-org/ieee/web/org/pubs/ieee-taxonomy.pdf}} (IEEE), it provides a standardized framework for organizing the Electrical and Electronics Engineering domain. This taxonomy is used to classify academic publications, research topics, and technical content within IEEE's publications and databases. Comprising approximately 5,600 topics and 24,000 relationships, the IEEE Taxonomy is manually curated and receives minor updates annually.

This paper presents the new methodology that will be used to update CSO in the following years and therefore will mostly focus on CSO.

\subsection{Ontology Generation}
A review of the current literature reveals various methods, both semi-automatic and fully automatic, for creating ontologies and taxonomies. The first step in developing an ontology involves identifying its core topics, and recent research is focused on streamlining this process using automated techniques. For example, BERT~\citep{devlin2019bert} has been employed for topic extraction, as demonstrated in~\citep{grootendorst2022bertopic}.

Historically, methods for ontology extraction have utilized NLP, clustering techniques, and statistical methods~\citep{10.1007/11428817_21,le-etal-2019-inferring}. One such method is Text2Onto~\citep{10.1007/11428817_21}, a framework designed to derive ontologies from document corpora. This framework uses NLP techniques to analyze sentence structures, identifying synonyms and hierarchical relationships such as subclasses and superclasses. It detects phrases like ``such as...'' and ``and other...'' to infer hierarchical connections between terms.

\cite{shen-etal-2018-web} adapted this method to create Fields of Study (FoS) for Microsoft Academic. Their approach combined manually curated concepts for the first two levels with topics automatically extracted from Wikidata. However, this taxonomy learning method mainly depends on Wikidata and overlooks the metadata from research papers.

In a similar vein, the OpenAlex team implemented a related approach~\citep{openalex,openalexpaper}. They incorporated the ASJC structure from Scopus and supplemented it with topics derived from papers through citation analysis.

Other approaches have explored the fusion of ontology learning with crowdsourcing methods, blending statistical analyses with user feedback~\citep{Wohlgenannt2012,Mortensen2013}. For instance,~\cite{Wohlgenannt2012} combined human assessment with computational methods by using crowdsourcing to evaluate an automatically generated ontology, aiming to dynamically validate extracted relationships.

Recently, there has been exploration within the community into using LLMs to create taxonomies, ontologies, and Knowledge Graphs (KGs)~\citep{allen2023knowledge}. For instance,~\cite{Chen2020ConstructingTF} proposed a method for generating taxonomies consisting of two modules: one module predicts parent-child relationships, while the other module consolidates these predictions into hierarchical trees. The parent-child prediction module assigns likelihood scores to potential parent-child pairs, forming a graph based on these relation scores. The tree reconciliation module approaches the task as a graph optimization problem to derive the maximum spanning tree from this graph. The model is trained on subtrees sampled from Wordnet and evaluated on distinct, non-overlapping Wordnet subtrees.

To our knowledge, there are currently no established methodologies using language models specifically for creating research topic ontologies.

The proposed work aims to pioneer the creation of a comprehensive taxonomy in the field of computer science. Our approach integrates advanced artificial intelligence techniques with extensive datasets sourced from CSO. Through this innovative blend of AI and data-driven insights, our goal is not only to refine existing taxonomies but also to establish a flexible framework that can adapt to the evolving landscape of computer science research.

\section{Background Data}
\label{sec:background}

In this section, we present the two main datasets used in this study. The first is the novel CSO-21K benchmark, which was employed to evaluate the relation classification component. The second is the Academia/Industry Dynamics Knowledge Graph (AIDA-KG)~\citep{angioni2021aida}, a large repository of research paper metadata. This dataset is used by our methodology to extract features related to the occurrence of research topics in scientific publications.


\subsection{The CSO-21K benchmark}
\label{subsec:gs}



The CSO-21K benchmark was developed with the goal of creating a large-scale dataset for research topic relationship classification. 
We formalize this task as a single-label, multi-class classification problem, where a pair of topics ($t_A$ and $t_B$) is assigned to one of the following categories:
\begin{itemize} \item \textit{supertopic}: $t_A$ is an ancestor of $t_B$, such as \textit{databases} being a broader area encompassing \textit{SQL}; \item \textit{subtopic}: $t_A$ is a descendant of $t_B$, for instance, \textit{deep learning} being a specialized topic under \textit{artificial intelligence}; \item \textit{same-as}: $t_A$ and $t_B$ represent alternative labels for the same topic, e.g., \textit{haptic interface} and \textit{haptic device}; \item \textit{other}: $t_A$ and $t_B$ do not fall into any of the previous relationships, such as the pair \textit{cryptocurrency} and \textit{pervasive computing}. \end{itemize}


In particular, since our goal was to develop a successor to Klink-2 for generating a new version of CSO, we built our gold standard by extracting the manually validated topics and relationships from CSO~\citep{salatino2018computer}.
The CSO, accessible through the CSO Portal\footnote{CSO Portal: \url{https://cso.kmi.open.ac.uk}}, is an evolving resource that undergoes continuous expert review, particularly in the context of specific scientometric studies (e.g., Software Engineering~\citep{osborne2019reducing}). We used the manually verified sections of CSO, which were reviewed and refined by researchers and domain experts, to construct the gold standard employed for training. 
While CSO adopts a slightly different ontological schema from the one described above, it can be easily mapped to the schema used in this work. Specifically, we mapped the CSO \textit{superTopicOf} relation to our \textit{superTopic} relation, and the CSO \textit{relatedEquivalent} relation to our \textit{same-as} relation.

We first extracted 5,796 manually verified \textit{superTopic} relations from the validated subset of CSO. By reversing these relations, we obtained an additional 5,796 \textit{subTopic} relations. We then selected 3,692 \textit{same-as} relations, again relying exclusively on the manually validated portion of CSO to ensure high accuracy. Finally, we generated 6,365 \textit{other} relations by randomly pairing topics, making sure to exclude any pair that matched a \textit{superTopic}, \textit{subTopic}, or \textit{same-as} relation found in the complete CSO ontology. Although this selection process was stochastic, we applied a conservative filtering strategy to eliminate any pairs that could plausibly be semantically related according to the broader CSO knowledge base. The number of \textit{other} pairs was chosen to ensure a balanced distribution across classes while still preserving a sufficient level of diversity in negative examples. 
The resulting dataset comprises 21,649 topic pairs, each annotated with one of the four relation types.

The dataset was then split into 15,154 triples for the training set (70\%), 2,166 triples for the validation set (10\%), and 4,239 triples for the test set (20\%). Importantly, we ensured that if a topic pair \textit(A, B) appears in any split (training, validation, or test), the reverse pair \textit(B, A) is included in the same split. This constraint prevents any pair and its inverse from appearing across different subsets, thereby avoiding potential data leakage and preserving the integrity of the evaluation.

\subsection{AIDA Knowledge Graph}
\label{subsec:AIDA}

The pipeline presented in the next section leverages a large repository of research papers to extract numerical features that characterize the relationship between two topics, such as their individual occurrences and their co-occurrence in scientific articles. 
As the data source, we used AIDA KG~\cite{angioni2021aida}, an extensive KG that covers 25 million publications and 8 million patents in the Computer Science domain. The papers in AIDA KG are associated with research topics from CSO, researcher profiles, and industrial sectors. 
This KG provides rich metadata for each document, including authors, affiliations, countries, venues, and relevant research topics.

AIDA KG was created through an automated pipeline that integrates data from OpenAlex, DBLP, Research Organization Registry (ROR), and DBpedia. AIDA KG encompasses eight entity types: papers, patents, authors, affiliations, journals, conferences, topics, and industrial sectors. These entities are linked by 22 relations, such as `hasAffiliation', which specifies author affiliations; `hasTopic', connecting documents to research topics from the Computer Science Ontology; `hasIndustrialSector', linking documents to economic activity domains; and `schema:creator', indicating a paper's author. The underlying ontological schema of AIDA KG, available at \url{https://w3id.org/aida#aidaschema}, conforms to the Resource Description Framework (RDF) standard, and the data can be accessed as a download or queried via a triplestore at \url{https://aida.kmi.open.ac.uk/sparql/}, under a CC BY 4.0 license.

\section{Methodology}
\label{sec:method}

This section presents a detailed description of the methodology for semi-automatic generation of research topic ontologies from a large-scale repository of research papers.


%

\begin{figure}
    \centering
    \includegraphics[width=\textwidth]{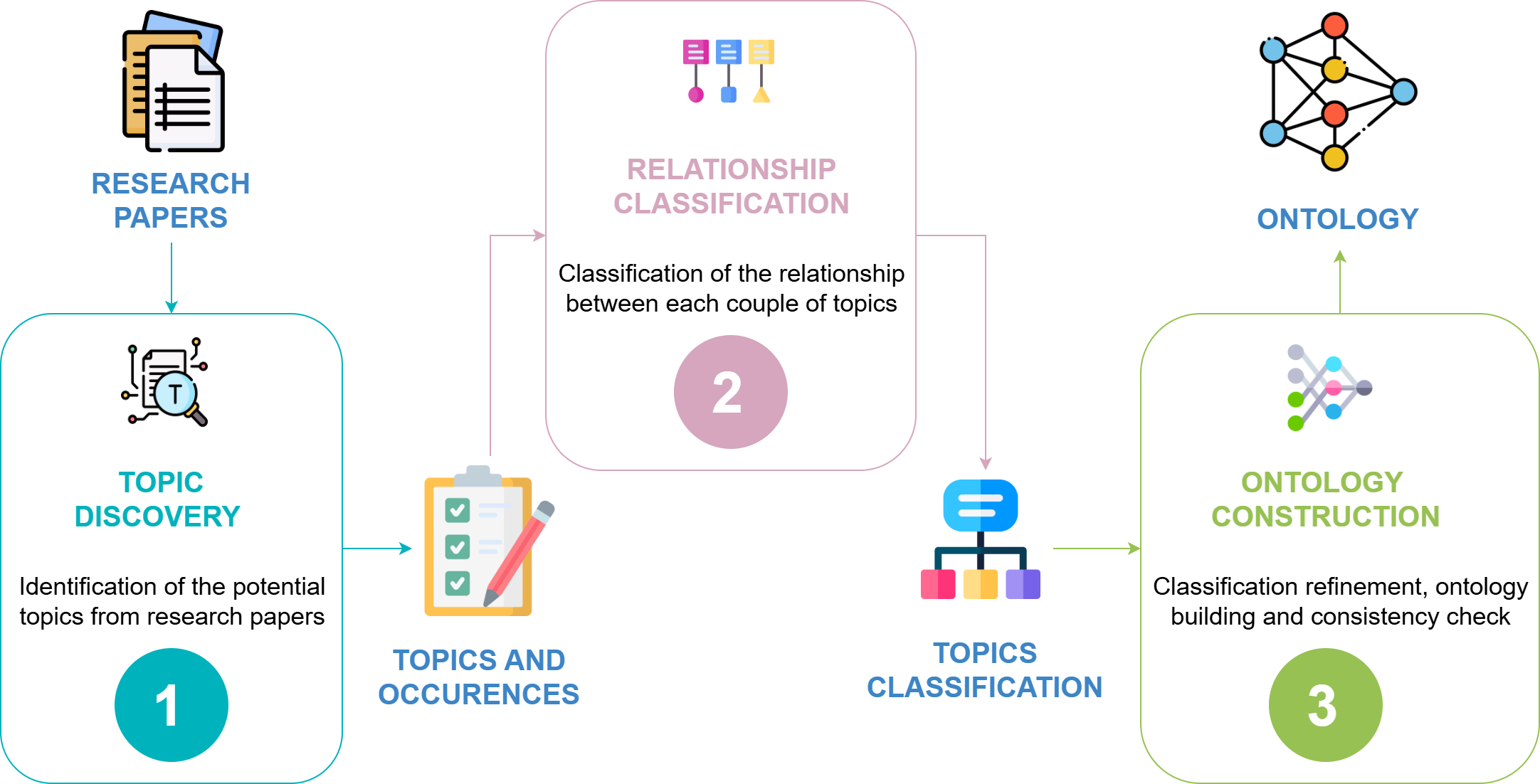}
    \caption{Overview of the pipeline.}
    \label{fig:pipeline_overview}
\end{figure}

\Cref{fig:pipeline_overview} illustrates the architecture of our approach which consists of three main steps:

\begin{itemize}
\item \textbf{Topic Discovery}. This step takes as input a set of research papers and identifies potential topics within their content. It outputs a list of the identified topics. Additionally, for each topic, this phase returns its \textit{occurrences}, defined as the total number of times the topic appears in the abstracts of the input research papers.
\item \textbf{Relationship Classification}. Given the topics and their occurrences extracted in the previous step, this phase generates pairs of topics and assigns a class that represents their relationship. It outputs a list of topic pairs along with their inferred relationship.
\item \textbf{Ontology Construction}. The final stage refines the topic pairs and their assigned classes to build the resulting ontology, which is then returned as output.
\end{itemize}

We describe the three stages in detail in the following sections.


\begin{figure}
    \centering
    \includegraphics[width=0.6\textwidth]{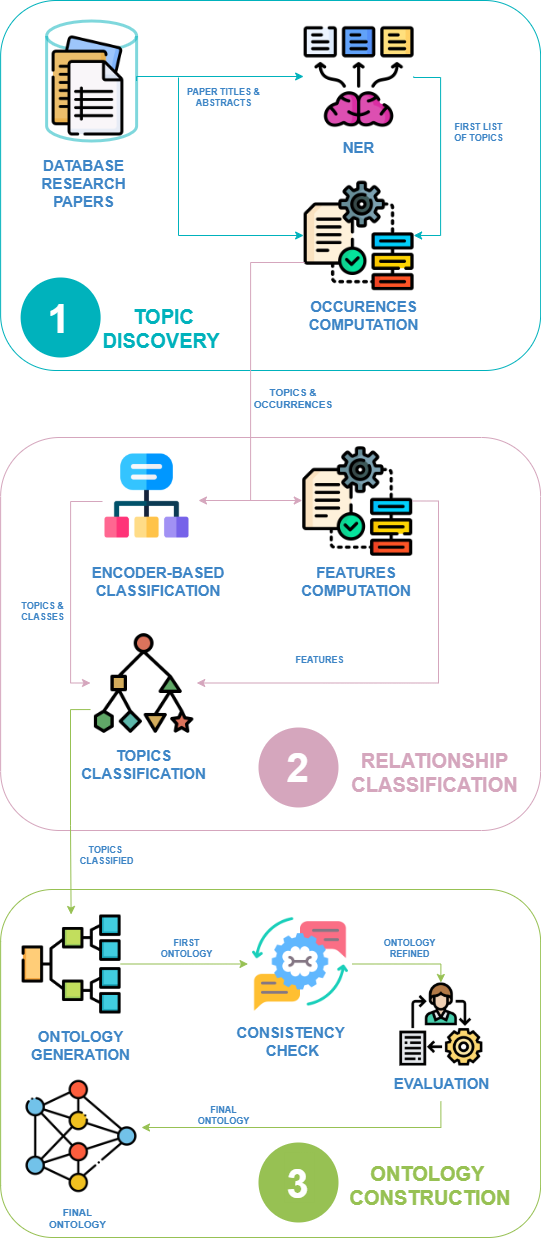}
    \caption{Detailed pipeline.}
    \label{fig:pipeline_detail}
\end{figure}

\subsection{Topic Discovery}
\label{subsec:topic_discovery}

The aim of this phase is to identify a new set of candidate topics in order to build an ontology that represents the scientific landscape of a given field. To this end, we use as input a collection of publication metadata, where each paper is described by its title, abstract, and year of publication. We then extract candidate topics from the text using a Named Entity Recognition (NER) method. Subsequently, for each topic, we compute its frequency in the dataset as well as its co-occurrence with other topics. This information is used in the subsequent phases.

After conducting preliminary tests with different types of textual sources, we decided to rely on the title and abstract in this step.
The combination of title and abstract provides a solid representation of the main topic of a paper: the title highlights the most important concept, while the abstract typically summarizes the motivation, methodology, and findings in a concise form, making it a particularly effective source for topic extraction.
This pair offers sufficient context to capture a broad range of scientific concepts, while remaining computationally manageable. 
We do not use keywords because their quality varies significantly depending on the publisher or the creator of the paper's metadata. Moreover, they are often already mentioned in the title or abstract, making them redundant. Finally, they do not necessarily correspond to scientific topics, as authors frequently include the names of technologies, software, or tools as keywords.
Similarly, we chose not to use the full text for three main reasons. First, many public repositories provide only the metadata of the paper and not the full text; when the full text is available, it is typically in PDF format and requires significant preprocessing. Second, processing such a large amount of text introduces scalability challenges. Finally, extracting topics from the full text often produces a noisy set, where the most relevant and useful topics are usually the same as those that can be extracted from the abstract. 
However, it is important to note that, apart from scalability considerations, our topic extraction pipeline can also be applied to additional textual sources. In this work, we focused on the title and abstract because they appear to be the most suitable for our primary use case, which is the creation of new versions and branches of CSO.

In terms of repositories, our method can be applied to most current publication metadata repositories, as all of them provide at least the title, abstract, and publication year. These include OpenAlex \cite{priem2022openalex, openalex, openalexpaper}, Scopus \cite{scopus}, CORE \cite{core}, Semantic Scholar \cite{semanticscholar}, CiteSeerX \cite{citeseerx}, OpenCitations \cite{opencitations}, and others.
In the current prototype, we used AIDA-KG, discussed in the background section, which is a KG of computer science papers and represents an enriched version of OpenAlex.

In the following, we describe how we 1) extract topics from the title and abstract, and 2) compute the occurrences of the selected topics.

\textbf{Extracting research topics from text.}
In order to identify the topics associated with a paper based on its abstract and title, we formulate the task as a NER problem.
NER is an information extraction method that assigns parts of a text to predefined categories. In our case, we classify the text into three distinct classes:
\begin{itemize}
\item \textbf{B-TOP}, assigned to the first token of a multi-token research topic;
\item \textbf{I-TOP}, assigned to each subsequent token that is part of the research topic but not the first;
\item \textbf{MISC}, assigned to each token that does not belong to any research topic.
\end{itemize}

To create the NER model, we fine-tune a pretrained BERT model \cite{devlin2019bert}, which typically performs very well on this task \cite{10724703}.

We generate a dataset for fine-tuning BERT by applying the CSO Classifier to a sample of 10,000 papers from the AIDA Knowledge Graph. The CSO Classifier~\cite{salatino2019b} is an approach that links portions of text to topics from the Computer Science Ontology (see Section~\ref{subsec:gs}). Unlike the model we aim to train, it is limited to the set of topics currently defined in CSO. However, we used it here solely to annotate all text segments that could be linked to existing topics, assigning them the \textbf{B-TOP} and \textbf{I-TOP} labels. This enabled us to leverage existing tools from CSO to produce a high-quality dataset for fine-tuning a system capable of recognising research topics within text.

During preprocessing, papers missing key attributes (such as title or abstract) or not written in English were discarded. The remaining text was then merged and segmented into individual sentences to better capture contextual information and simplify annotation.  
The resulting dataset contains approximately 63,000 sentences, which were randomly split into training (70\%), validation (15\%), and test (15\%) subsets. Tokens corresponding to identified topics were labeled with the appropriate \textbf{B-TOP} or \textbf{I-TOP} tags, while all other tokens were labeled as \textbf{MISC}.

To produce the best model, we performed a grid search by varying the number of epochs [1, 2, 3], warm-up steps [10, 20, 40, 50, 100, 300, 400, 500], training batch sizes [4, 8, 10], and learning rates [0.001, 0.0001, 0.00001]. The optimal configuration used 2 epochs, 50 warm-up steps, a batch size of 10, and a learning rate of 0.0001.

Since the NER approach can sometimes generate overly generic topics, we introduce a post-processing step to filter out common words that occur with high frequency in the repository. In the prototype, we integrate this filter with an allow-list, consisting of topics already present in known ontologies, and a deny-list of undesirable words, which we construct through extensive methodological experimentation.


%

To assess the quality of the model in extracting topics, we performed a manual validation. We randomly selected 150 concepts extracted by the model from a set of papers and presented the list to three experts, who independently evaluated whether each concept represented a valid research topic. 
Each expert provided a binary judgment (1 = valid topic, 0 = not a valid topic), and their responses were aggregated to construct a reference annotation. We considered a concept to be valid if it was judged positively by at least two out of three experts (i.e., a majority consensus). According to this criterion, 128 out of 150 concepts (84.67\%) were confirmed as valid research topics by the annotators. The remaining concepts, although generally relevant, were typically considered unsuitable because they were either too generic or overly specific, often referring to particular technologies or methodologies. 
This result indicates a high level of precision in the model's ability to identify meaningful research topics. 

\textbf{Computing topics occurrences.}
Finally, we compute the occurrences and co-occurrences of the candidate topics extracted with the NER model. Specifically, given a topic \textit{$t_A$}, we calculate its frequency in the abstracts of papers published over the past ten years, as well as its co-occurrence with all other candidate topics.
This information will be used as a feature in the next phase, which focuses on identifying relations between topics.

%

\subsection{Relationship Classification}
\label{subsec:relationship_classification}

In this phase, we aim to infer the relations between each pair of candidate topics. 

In most practical cases involving CSO, the process is sufficiently scalable to allow exploration of the relationships between each topic and all others. If scalability becomes an issue, an alternative approach is to compare only those topics whose cosine similarity exceeds a given threshold.



As discussed in Section 3.1, we formalize this task as a single-label, multi-class classification problem, where each pair of topics is assigned to one of the following four categories: \textit{supertopic} ($t_A$ is an ancestor of $t_B$), \textit{subtopic} ($t_A$ is a descendant of $t_B$), \textit{same-as} ($t_A$ and $t_B$ are alternative labels for the same topic), and \textit{other}.

This step can naturally be performed by processing the labels of the topics with language models fine-tuned for this task. However, our experiments suggest that it is also very beneficial to include additional features derived from the actual usage of the topics within the scientific literature. We therefore adopt a hybrid methodology that combines an encoder-based language model with a classifier capable of processing additional numeric features. Section~\ref{sec:evaluation} reports the evaluation, which demonstrates that this solution is superior to alternative ap\-proa\-ches such as fine-tuning BERT or GPT-4o mini. 

In the following, we first describe the steps performed with the language model, then introduce the numerical features, and finally present the overall classification process, which integrates the output of the language model with these features.

First, we perform the classification task by giving in input the surface form of the pair of topic to a SciBERT model~\cite{Beltagy2019SciBERT} fine tuned on the training set of the CSO-21k dataset, presented in Section~\ref{subsec:gs}.

This produces a first approximation of the classification that we then enrich with additional features about topic usage in the literature. The set of features were chosen based on those used by Klink-2, the previous approach for generating CSO~\citep{osborne2015klink,10.1007/11428817_21}, as well as on features commonly adopted in traditional taxonomy generation methods~\citep{sanderson1999deriving}.
For each pair of topics (\textit{$t_A$}, \textit{$t_B$}), we consider four features: 1) \textit{occA}, the frequency of $t_A$ in paper abstracts; 2) \textit{occB}, the frequency of $t_B$; 3) \textit{cooccurrenceAB}, the frequency of their co-occurrence; and 4) their \textit{subsumption}, which measures the overlap between co-occurring topics and is calculated as $\frac{cooccurrenceAB}{occA} - \frac{cooccurrenceAB}{occB}$.

In order to integrate this set of features with the output of the SciBERT classifier, we adopt a \textit{Random Forest} classifier. The classifier takes as input the four numeric features, along with a one-hot encoding of the class predicted by SciBERT, and produces one of the four categories. As demonstrated in~\Cref{sec:evaluation}, this step further refines the categories suggested by the language model by incorporating knowledge about the usage of the topics in the literature, resulting in more robust and accurate results.

\subsection{Ontology Construction}
\label{subsec:ontology_construction}

The final stage of our methodology takes the previously generated list of topic pairs and their associated relations, refines them, and constructs the final ontology.

It is important to note that the process is designed to be semi-automated, with the support and guidance of human experts, because creating such an ontology often involves nuances that cannot be fully captured by automation. Nevertheless, the proposed methodology automates the majority of the time-consuming tasks, enabling humans to concentrate on reviewing and refining the final taxonomy structure. This greatly reduces the overall effort required for the process.
In the following, we will describe the various steps of this phase.

\textbf{Consistency check.}

As a first step, to avoid semantic inconsistencies, we discard a concept pair when the relation assigned to the pair $(A, B)$ is semantically incompatible with the relation assigned to its inverse pair $(B, A)$.
For instance, if the pair ('machine learning', 'random forest') is assigned the class \textit{super-concept}, then the inverse pair ('random forest', 'machine learning') must be assigned the class \textit{sub-concept}. Assigning any other class would create a semantic contradiction, since these classes are mutually exclusive. Similarly, if the pair ('knowledge\,graph', 'block\-chain') is labelled as \textit{other}, the inverse pair ('blockchain', 'knowledge graph') cannot be assigned any class other than \textit{other}. 
Any such inconsistencies undermine the integrity of the classification. Therefore, to maintain overall consistency, all concept pairs involved in these conflicts are discarded and excluded from further analysis.




%

\textbf{Taxonomy generation.}
The process begins by defining a root topic, denoted as \textit{root concept} (e.g., \textit{cyber security}), and then expands it by following the \textit{same as} and \textit{subtopic} relations. It then performs a depth-first search by selecting one of its subtopics and expanding it in the same way. This procedure continues iteratively, with each subsequent subtopic being expanded until either no new subtopics are found or a maximum number of iterations is reached.
The resulting taxonomy of topics provides a first approximation of the ontology and is stored as a JSON file in the following format: \texttt{\{ `topic': \{ `supertopic': [], `subtopic': [], `same-as': [] \} \}}.  
However, it requires further refinement before it can be used effectively, for example, by detecting and breaking loops that could otherwise result in an invalid or unmanageable taxonomy. 


\textbf{Validation of same-as.}
For each concept in the taxonomy, the \textit{same-as} field, which contains potential synonyms, abbreviations, or terminological variants, is analyzed. A validation procedure is applied to ensure that only semantically robust equivalences are retained. This procedure relies on two criteria:
\begin{itemize}
\item \textit{acronym matching}: a candidate acronym must exactly match the initial letters of the corresponding topic (e.g., "NLP" for "Natural Language Processing") to be considered valid;
\item \textit{semantic similarity} based on embedding models: each pair is evaluated using vector-based semantic similarity measures, and only pairs with a similarity above a predefined threshold are accepted.
\end{itemize}
All pairs that do not satisfy these criteria can be problematic in practice. Therefore, they need to be analyzed and vetted by a human expert; if this is not possible, the underlying \textit{same-as} relation is discarded.

\textbf{Alternative label definition.}
Concepts mutually connected via same-as relations are grouped into coherent clusters, each representing a potential unified semantic entity that can be interpreted as a research topic. For each cluster, one concept is designated as the \textit{main label} of the cluster. In our current prototype, the concept with the highest number of occurrences in the source data is selected. This choice is motivated by the assumption that a higher number of occurrences reflects greater prevalence in the scientific literature, making it a more recognizable and representative label for the corresponding cluster. All remaining concepts in the cluster are treated as alternative labels, capturing the terminological variation of the topic.

At this stage, a novel representation of the taxonomy structure is generated, where the main label represents each ontology node in the following format:
\texttt{\{ `topic': \{`main\_label': `topic', `supertopic': [], `subtopic': [], `alternative-label': []\}\}}.

\textbf{Breaking loops.}
Finally, a structural validation step is performed to remove potential cycles in the hierarchical relations. A depth-first search algorithm is applied to the graph to detect cyclic paths within the supertopic–subtopic links. When a cycle is detected, the system evaluates the involved edges and removes the least informative one. Specifically, it removes the edge with the lower co-occurrence between the two topics. The updated structure is then analyzed again to ensure that no further cycles remain. Once a valid acyclic structure is confirmed, the ontology is saved as the final output.

\textbf{Human evaluation.}
Finally, a manual review is conducted by domain experts. This step is essential to ensure that the ontology complies with formal semantic and structural criteria, while also capturing accurate and meaningful relationships from a domain-specific perspective.
During this phase, an expert can:
\begin{itemize}
\item \textit{add new relationships} between topics;
\item \textit{remove incorrect relationships} that are deemed invalid;
\item \textit{discard topics or alternative labels} that, for any reason, are considered inappropriate.
\end{itemize}
The resulting validated structure combines the scalability of automated processing with the specialist oversight that ensures the ontology is both robust and representative of the scientific landscape.

\section{Experimental evaluation}
\label{sec:evaluation}

In this section, we present the evaluation of the relation classification task, which is the core component of our ontology generation approach. Specifically, we compare the hybrid classifier employed in our methodology in Section~\ref{subsec:relationship_classification} with several alternative baselines, using the CSO-21k benchmark. 

Specifically, we address the task described in detail in Section 3.1, which involves classifying a pair of topics according to their relationship, encoded into four categories: (\textit{supertopic}, \textit{subtopic}, \textit{same-as}, and \textit{other}).
All the tested methods were trained on the training and validation sets of the CSO-21k dataset, which consist of 15,154 and 2,166 topic pairs, respectively.
They were then evaluated on the test set (4{,}239 topic pairs) using standard metrics for text classification: accuracy, precision, recall, and F-measure.

In the following, we outline the classification methods and report the results.




\subsection{Classification methods}
\label{subsec:methods}

We evaluated 14 alternative solutions on CSO-21K. First, we consider the following nine classification methods, which are based on different combinations of features and encoders:

\begin{itemize}
\item \textbf{Fine-tuned SciBERT (FT SB)}. This method, also used in~\cite{MottaSOPPRR24}, corresponds to the first part of our approach and does not incorporate the additional features from the literature.  

\item \textbf{Classifier based on aggregate features (AF)}. This approach uses as input the four features described in Section~4.2. We implemented two versions of this classifier, based on \textit{Random Forest} (AF + RF) and \textit{Gradient Boosting} (AF + GB).  

\item \textbf{Classifier based on yearly features (YF)}. This method uses the same features as AF, but broken down by year. This design allows us to capture temporal trends that may be obscured when considering only the overall statistics. We computed these features from papers published over the last ten years, treating each year individually. As a result, we derived 40 distinct temporal features. We implemented two versions of this classifier, based on \textit{Random Forest} (YF + RF) and \textit{Gradient Boosting} (YF + GB).

\item \textbf{Classifier based on aggregate features + SciBERT prediction (AF + SB)}. This method corresponds to the final version of our approach, as described in Section~4.2, and combines the aggregate features with the SciBERT prediction. As before, we tested both \textit{Random Forest} (AF + SB + RF) and \textit{Gradient Boosting} (AF + SB + GB) as classifiers.  

\item \textbf{Classifier based on yearly features + SciBERT prediction (YF + SB)}. In this variant, we augment the yearly features with additional ones derived from the SciBERT prediction, following the same rationale as in AF + SB. As before, we evaluated this method using both classifiers (YF + SB + RF, YF + SB + GB).  
\end{itemize}

Furthermore, we compared these classifiers with a range of LLM-based methods under three main configurations: zero-shot learning, Chain of Thought (CoT), and fine-tuning. 

For zero-shot learning, we adopted the reasoning techniques proposed in~\cite{Aggarwal} and~\cite{aggarwal2026large}. This approach uses a carefully designed prompt that explicitly defines the semantics of the four relationships and asks the LLM to classify a given pair of topics accordingly. We refer to this method as \textbf{Standard (STD)}. 
For the \textbf{Chain of Thought (CoT)} configuration, we applied the technique from~\cite{aggarwal2026large}. In this method, the model is prompted to reason step by step: first providing precise definitions of the two topics, then discussing their usage and potential relations, and finally determining their actual relationship. We used \textit{GPT-4 Turbo} and \textit{GPT-3.5} for both the STD and CoT configurations. 

Finally, we fine-tuned a \textit{GPT-4o Mini} model (\textbf{GPT4-m. FT}). We chose this model because it is a standard OpenAI option for fine-tuning, providing a good balance between performance and cost. The fine-tuning was performed using the official OpenAI platform, with a batch size of 20 and a learning rate multiplier of 1.8.

\subsection{Experimental results}
Table~\ref{tab:acc-f1} reports the performance of the 14 alternative approaches in terms of accuracy and F1.




\begin{table}[b!]
  \centering
  \normalsize
  \caption{Experimental results in terms of accuracy and F-measure. Abbreviations: supert. = supertopic; subt. = subtopic; FT = Fine tuned; SB = SciBERT; AF = Aggregate features; YF = Year features; RF = Random forest; GB = Gradient boosting; CoT = Chain of Thought; STD = Standard; GPT-4.t = GPT-4.turbo; GPT-4.m = GPT-4.mini. The best results for each metrics are represented in bold.}
\begin{tabular}{l|c|r|r|r|r|r}
\cmidrule[\heavyrulewidth]{3-7}
\multicolumn{1}{c}{} & \multicolumn{1}{c}{} & \multicolumn{5}{|c}{\textbf{F-measure}}                                                   \\ \midrule
\textbf{Method} & \textbf{Accuracy}                               & \textbf{supert.} & \textbf{subt.} & \textbf{same-as} & \textbf{other} & \textbf{overall} \\ \midrule
FT SB                                                & 0.950                & 0.944            & 0.945          & \textbf{0.950}   & 0.958          & 0.949            \\
AF+RF                                                & 0.581                & 0.472            & 0.477          & 0.526            & 0.734          & 0.552            \\
AF+GB                                                & 0.772                & 0.750            & 0.756          & 0.675            & 0.863          & 0.761            \\
YF+RF                                                & 0.581                & 0.472            & 0.477          & 0.526            & 0.734          & 0.552            \\
YF+GB                                                & 0.830                & 0.824            & 0.824          & 0.716            & 0.901          & 0.816            \\
AF+SB+RF                                             & \textbf{0.952}       & 0.945            & 0.945          & \textbf{0.950}   & 0.964          & \textbf{0.951}   \\
AF+SB+GB                                             & \textbf{0.952}       & 0.944            & 0.946          & \textbf{0.950}   & \textbf{0.966} & \textbf{0.951}   \\
YF+SB+RF                                             & \textbf{0.952}       & \textbf{0.946}   & \textbf{0.947} & 0.948            & 0.964          & \textbf{0.951}   \\
YF+SB+GB                                             & \textbf{0.952}       & 0.945            & 0.946          & 0.948            & 0.965          & \textbf{0.951}   \\
GPT4-t. CoT                                          & 0.698                & 0.634            & 0.643          & 0.671            & 0.803          & 0.688            \\
GPT4-t. STD                                          & 0.737                & 0.667            & 0.660          & 0.786            & 0.822          & 0.734            \\
GPT3.5 CoT                                           & 0.380                & 0.430            & 0.441          & 0.403            & 0.120          & 0.348            \\
GPT3.5 STD                                           & 0.297                & 0.428            & 0.225          & 0.011            & 0.155          & 0.205            \\
GPT4-m. FT                                           & 0.934                & 0.922            & 0.929          & 0.906            & 0.965          & 0.930  \\
\bottomrule

\end{tabular}
  \label{tab:acc-f1}%
\end{table}%

The hybrid method introduced in this paper achieved the highest accuracy (0.952) and F1 score (0.951), outperforming both the fine-tuned SciBERT model and all the LLMs, including the fine-tuned GPT4-mini.   
Interestingly, some alternative configurations of the proposed approach, which used features split by year and the GB instead of the RGF, produced almost identical results. This finding thus suggests that the key factor is the integration of information from the language model with the features derived from the literature. 
Since splitting the features by year did not improve performance and introduced additional complexity, we adopted the aggregated version of the features in our prototype.

Notably, while the fine-tuned GPT4-mini model achieves robust results (0.934 F1), the non-fine-tuned LLM configurations yield rather mediocre performance. The best among them is GPT4-t STD, with an F1 score of  0.734. This suggests that versions of LLMs not fine-tuned for this specific task do not perform well, even when using advanced prompt optimization strategies such as the one proposed in~\cite{aggarwal2026large}.

Finally, the four methods that rely solely on Year or Aggregate features, without incorporating SciBERT predictions, achieved significantly lower accuracy scores, ranging from 0.581 (AF + RF and YF + RF) to 0.830 (YF + GB). Notably, when using only these features, the GB classifier significantly outperforms the RF classifier.

\section{Case Study}
\label{sec:casestudy}

To demonstrate the effectiveness of the proposed approach, we present a use case involving its application to extending CSO in the field of cybersecurity. It is often necessary to extend particular branches of CSO either to support scientometric analyses or to respond to requests from communities that feel their branch is underdeveloped. This is the case for cybersecurity, which in the previous version of CSO was represented only by 1 subtopic with 1 descendant, resulting in a coarse-grained and limited representation of articles and events in this field.

We therefore applied the methodology described in \Cref{sec:method}, using \textit{cybersecurity} as the root and selecting as input a sample of 15 million scientific articles published in the past two years in the field of computer science, sourced from AIDA KG. 
The titles and abstracts of these papers were extracted and passed to the first step of our pipeline. This initial module identified approximately 50K distinct scientific concepts across the entire field of computer science.

Focusing on cybersecurity, we then selected a subset of 500 topics that showed the highest co-occurrence frequency with the term \textit{cybersecurity}. This ensured that the selected concepts were semantically related to the root and relevant to the area under investigation.

To streamline the subsequent stages of the pipeline, the selected topics underwent a review process conducted by domain experts, who removed those deemed too distant from or irrelevant to the cybersecurity domain. This filtering process yielded a final set of 37 cybersecurity-related topics, which were subsequently forwarded to the second stage of the pipeline, responsible for inferring semantic relations among them.

The system automatically generated the ontology structure starting from the topic \textit{cybersecurity}. The resulting ontology was subsequently subjected to a qualitative validation phase conducted by domain experts, who refined the structure and corrected relationships that were considered inaccurate or irrelevant in order to enhance its conceptual coherence and relevance to the domain. 


\begin{figure}
    \centering
    \includegraphics[width=1\linewidth, scale=1]{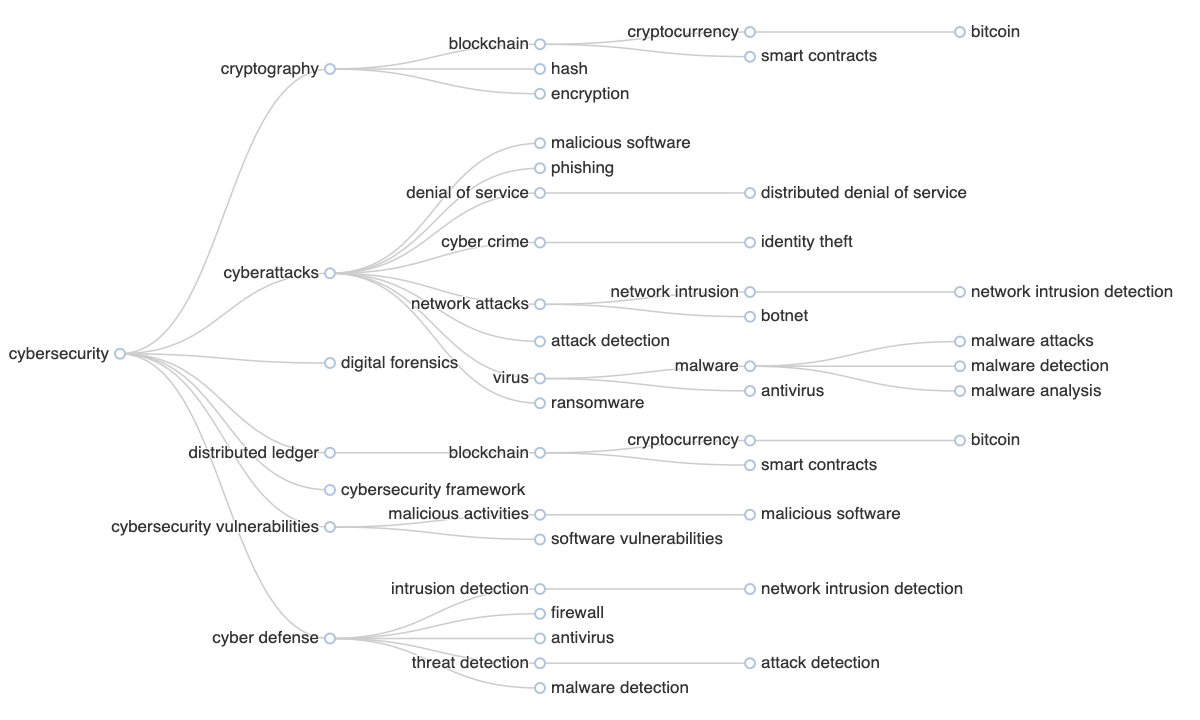}
    \caption{Cybersecurity ontology.}
    \label{fig:cybersecurity_ontology}
\end{figure}

This resulted in the creation of a \textit{cybersecurity} branch in the ontology, covering 37 topics distributed across 4 levels. The ontology is represented in  \Cref{fig:cybersecurity_ontology}.  At the first level, directly subordinate to the root, lie 7 core thematic areas that constitute foundational pillars in the study and practice of cybersecurity. These include \textit{cryptography}, \textit{digital forensics}, \textit{cybersecurity vulnerabilities}, \textit{distributed ledger}, \textit{cyberattacks}, \textit{cybersecurity fra\-mework}, and \textit{cyber defense}. These domains are treated as primary specializations and are semantically distinguished yet interconnected through underlying dependencies. 

The second level introduces subdomains that refine the semantic scope of their respective parent concepts. For example, under \textit{cryptography}, nodes such as \textit{encryption}, \textit{hash}, and \textit{blockchain} emerge, representing well-established research topics in the study of secure information and data integrity. Similarly, the \textit{cyber defense} branch expands into topics like \textit{threat detection}, \textit{malware detection}, \textit{antivirus}, \textit{intrusion detection} and \textit{firewall}, indicating central research areas employed in that field. 

The third and fourth levels contain more specific topics, including distinct threat models, attack strategies or defence mechanisms. Examples within these levels include  \textit{malware analysis}, \textit{malware attacks}, \textit{identity theft}, and \textit{distributed denial of service}, which correspond to concrete manifestations and challenges within the field of \textit{cybersecurity}.


The new cybersecurity branch was integrated into the overall structure of the CSO by importing the corresponding triples, after verifying that they did not introduce any inconsistencies. Following this extension, the \textit{cybersecurity} topic now includes 7 subtopics and 611 descendant topics, offering a substantially richer and more detailed representation. The number of descendant topics exceeds that of the newly added topics because some topics (e.g., cryptography) were already present in the CSO and had their own subtopics, which were also incorporated into the hierarchy. 
The \textit{cybersecurity} topic in CSO is available for consultation at the following URL: \url{https://cso.kmi.open.ac.uk/topics/cybersecurity}.

The creation of this ontology only required a single session with a domain expert. In contrast, the development of similar ontologies typically takes months or even years and involves multiple iterative sessions with domain experts. Beyond the time investment, such a manual process also entails substantial financial costs. 
Moreover, a manual approach to building this ontology would remain prone to inconsistencies and omissions. This is because domain experts often hold different perspectives, and the choice of different experts can lead to different ontological representations. In addition, experts tend to form a mental model of the domain that does not necessarily evolve to comprehensively reflect new research directions. In contrast, the methodology presented in this paper integrates objective, bottom-up information about emerging novel topics and current trends, derived from the millions of papers processed by our pipeline.  This produces a more complete representation in which emerging areas, which may not yet be widely recognized as important by all experts, are included if they have generated a substantial body of research in the venue under consideration.

In conclusion, the methodology proposed in this paper significantly reduces both the time and cost required for the task, while producing a more comprehensive representation of the various areas. This fine-grained representation of all novel topics is essential for enhancing downstream services, such as AI applications for analyzing the literature, recommendation systems, trend detection, and knowledge discovery. As a result, our methodology also contributes to more effective and efficient scholarly communication.



\section{Conclusions}
\label{sec:conclusions}

This paper presented Sci-OG, a novel methodology for the semi-automatic generation of taxonomies of research topics. The methodology has been implemented in the pipeline used to produce the Computer Science Ontology and has already been applied to extend the ontology, as demonstrated in the use case on cybersecurity. 
A crucial component of this system is the classifier for identifying relationships between pairs of topics, which combines an encoder-based language model fine-tuned on high-quality data with features describing topic occurrence in the scientific literature.
The evaluation on \textit{CSO-21K}, a novel dataset covering 21,649 relationships, shows that this solution achieves excellent performance and outperforms several alternative methods, including LLM-based baselines using GPT-4. The hybrid method proposed in this paper attained the highest accuracy (0.952) and F1 score (0.951), surpassing both the fine-tuned SciBERT model and all the LLMs, including the fine-tuned GPT4-mini.

This research advances the state of the art by proposing both a general methodology for ontology generation and a specific approach for relation classification. It also contributes to the development of AI-driven tools for literature management and exploration by providing a scalable solution for understanding and categorizing the rapidly growing body of scientific work.

Future work will progress along two main directions. First, we aim to develop a system that leverages LLMs to further support human experts in the manual steps of refining and validating the ontology. This system will be designed to assist experts by suggesting improvements, detecting inconsistencies, and streamlining the validation process. Second, we plan to extend our research to additional domains, including Engineering, Materials Science, and Biomedicine, in order to evaluate the generalizability and effectiveness of our approach across diverse scientific fields.




\section*{Acknowledgements}
We acknowledge financial support under the National Recovery and Resilience Plan (NRRP), Mission 4 Component 2 Investment 1.5 - Call for tender No.3277 published on December 30, 2021 by the Italian Ministry of University and Research (MUR) funded by the European Union – NextGenerationEU. Project Code ECS0000038 – Project Title eINS Ecosystem of Innovation for Next Generation Sardinia – CUP F53C22000430001- Grant Assignment Decree No. 1056 adopted on June 23, 2022 by the Italian Ministry of University and Research (MUR).

Alessia Pisu acknowledges MUR and EU-FSE for financial support of the PON Research and Innovation 2014-2020 (D.M. 1061/2021). She also acknowledges financial support under the National Recovery and Resilience Plan (NRRP), Mission 4 Component 2 Investment 1.3 - Call for tender No.341 published on March 15, 2022 by the Italian Ministry of University and Research (MUR), funded by the European Union - NextGenerationEU (Project Code PE0000013 – Project FAIR: Future Artificial Intelligent Research – CUP J53C22003010006), through a cascade call (Project ANSWER - innovAtive human-iN-the-loop-baSed knowledge undERstanding, CUP J23C24000080007).

Livio Pompianu acknowledges financial support from the project "FIATLUCS", funded by the PNRR RAISE Liguria, Spoke 01, CUP: F23C24000240006.

\bibliography{biblio}
\end{document}